\documentclass[epj]{svjour}
\usepackage{graphicx}
\usepackage{mathtext}
\usepackage{amsmath,amssymb}

\begin{document}

\title{Existence of the passage to the limit of inviscid fluid}

\author{Denis S.\ Goldobin}
\authorrunning{D.~S.~Goldobin}

\institute{Institute of Continuous Media Mechanics, UB RAS,
 Perm 614013, Russia
\and Department of Theoretical Physics, Perm State University, Perm 614990, Russia
}
\date{\today}

\abstract{
In the dynamics of viscous fluid, the case of vanishing kinematic viscosity is actually equivalent to the Reynolds number tending to infinity. Hence, in the limit of vanishing viscosity the fluid flow is essentially turbulent. On the other hand, the Euler equation, which is conventionally adopted for description of flow of inviscid fluid, does not possess proper turbulent behaviour. This raises the question of the existence of the passage to the limit of inviscid fluid for real low-viscosity fluids. To address this question, one should employ the theory of turbulent boundary layer near an inflexible boundary (e.g., rigid wall). On the basis of this theory, one can see how the solutions to the Euler equation become relevant for the description of flow of low-viscosity fluids, and obtain the small parameter quantifying accuracy of this description for real fluids.
%
%
}

\maketitle

\section{Introduction}
The flow of viscous incompressible fluid is governed by the Navier--Stokes equation,
\begin{equation}
\rho\left(\frac{\partial\vec{v}}{\partial t}+(\vec{v}\cdot\nabla)\vec{v}\right)
 =-\nabla p+\eta\mathrm{\Delta}\vec{v}+\rho\vec{g}\,,
\label{eq-01}
\end{equation}
and the continuity equation
\begin{equation}
\nabla\cdot\vec{v}=0\,,
\label{eq-02}
\end{equation}
Here we adopt conventional notations: $\rho$ is the fluid density, $\vec{v}$ is the flow velocity field, $p$ is pressure, $\eta$ is the dynamic viscosity coefficient, $\vec{g}$ is the specific mass force (if the only external mass force is the gravity, $\vec{g}$ is the gravity acceleration). A self-content mathematical description of the problem requires one to specify boundary conditions, and the order of these conditions must be consistent with the order of equations in the bulk. For the case of rigid boundary, these conditions are typically the no-slip conditions; for the case of nondeformable shear-stress-free boundary, the velocity component orthogonal to the boundary and the shear stress vanish, {\it etc}.

For the case of inviscid fluid, the Navier--Stokes equation turns into the Euler equation:
\begin{equation}
\rho\left(\frac{\partial\vec{v}}{\partial t}+(\vec{v}\cdot\nabla)\vec{v}\right)
 =-\nabla p+\rho\vec{g}\,,
\label{eq-03}
\end{equation}
As the order of eq.~(\ref{eq-03}) with respect to spatial derivatives is decreased compared to eq.~(\ref{eq-01}), the order of the boundary conditions, required for a self-content mathematical description, is decreased as well. In particular, the condition of no-slip of the flow along the boundary disappears; the remaining boundary condition is the condition that the boundary is impermeable.

However, the question of existence of the limiting case transition from the Navier--Stokes equation to the Euler equation arises. Indeed, where kinematic viscosity $\nu=\eta/\rho$ tends to zero, the Reynolds number $\mathrm{Re}=u_{\ast}L/\nu$ tends to infinity (here $u_\ast$ is the characteristic flow velocity and $L$ is the characteristic spatial scale of the system). As the Reynolds number tends to infinity, a developed turbulence sets-up in the system. Thus, small viscosity for real systems means generally the transition to essentially turbulent flow regimes, while the real turbulence
is inherent to the Navier--Stokes equation, but not the Euler equation, which describes the flow of inviscid fluids.

Generally, one can pose two distinct mathematical problems: the limiting case transition to inviscid fluid for (i)~laminar flow in Navier--Stokes equation and (ii)~turbulent flow. For the first case, one has a widespread problem of a small coefficient for the highest-order derivative in equations; for PDEs with such a small parameter the formation of thin boundary layers in time or space is typical. Beyond these layers the equations for vanishing parameter are valid, and one has to address the problems of characterization of the boundary layers and the derivation of the effective boundary conditions for the limiting equations in the bulk. These and associated problems ({\it e.g.}, the one of the uniqueness of solution) have a straightforward formulation and can be mathematically rigorously addressed. These problems for the limit of inviscid fluid have been extensively studied in the literature and significant advance has been made in proving the existence of the limiting case and characterization of its convergence for diverse situations ({\it e.g.}, see~\cite{Swann-1971,Kato-1972,Masmoudi-2007,Sammartino-Caflisch-1998a,Sammartino-Caflisch-1998b,Iftimie-Planas-2006,Constantin-Kukavica-Vicol-2015} and detailed review~\cite{Maekawa-Mazzucato-2016}). However, while for many physical processes (molecular diffusion, heat conductance, {\it etc.}) it is sufficient to consider the limiting case of vanishing coefficient of the highest-order derivative for ``laminar'' patterns, the Navier--Stokes equation can yield turbulence, for which the results of the consideration for ``laminar'' cases are not applicable. In particular, the effective turbulent viscosity appears and its value in the bulk (away from the boundaries) is not necessarily a small parameter. To the author's knowledge, the issue of the existence of the passage to the limit of inviscid fluid for the turbulent case did not receive attention comparable to the laminar case and remains unresolved.

The answer to the question of existence of this limiting case transition requires understanding of properties of turbulent currents in boundary layers near impermeable walls (rigid boundary or liquid--liquid interface). On the basis of the theory of the turbulent boundary layer, we briefly recall below, one can see how the solutions to the Euler equation may represent real fluid flows at low viscosity, and to estimate the accuracy of the representation of real flows by these solutions. The latter issue is a nontrivial one, because the characteristic small parameter of the system is determined by its physical parameters in a nonobvious way.

\begin{figure}[b]
\center{
 \includegraphics[width=0.24\textwidth]%
 {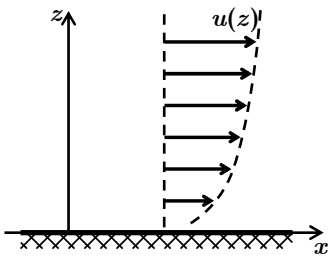}
}
\caption{Average flow in the turbulent boundary layer near a flat rigid wall}
\label{fig1}
\end{figure}

\section{Turbulent boundary layer}\label{sec2}
Since the consideration and derivations in the following sections heavily rely on the theory of the turbulent boundary layer~\cite{Karman,Prandtl}, we revoke here not only its results but also principal points. This will allow assessing the limits of applicability of the theory results for specific situations. A detailed consideration can be found in~\cite{Schlichting,Landau}.

Let us consider the flow of viscous fluid in a half-space near a flat rigid boundary. It is convenient to choose the boundary as the $(x,y)$-plane with the $x$-axis oriented along the average flow and the $z$-axis to be perpendicular to it (see fig.~\ref{fig1}). In a steady state (statistically stationary state), such a flow requires a spatially uniform shear stress $\sigma_{xz}$, which is related to the momentum flux towards the boundary, and the average flow $\vec{u}(\vec{r})$ itself must be a shear one; $\vec{u}(\vec{r})=\{u(z),0,0\}$.

In the turbulent case, the macroscopic average current is controlled by the effective turbulent viscosity. The turbulent viscosity is related to irregular pulsations of the velocity field which perform the momentum transport; for the case we consider, they yield the shear stress $\sigma_{xz}$. The uniform momentum flux performed by pulsations is determined by the intensity of these pulsations;
\begin{equation}
\sigma_{xz}\equiv\rho v_\ast^2\,,
\label{eq-04}
\end{equation}
where $v_\ast$ is the characteristic value of turbulent pulsations. Notice, equation~(\ref{eq-04}) is not an expression for the relation between $\sigma_{xz}$ and $v_\ast$, but serves as a definition of the introduced parameter $v_\ast$. With this definition, one can consider the problem in terms of the governing parameter $v_\ast$, in place of the original governing parameter $\sigma_{xz}$.

For the process of transfer of the momentum by turbulent pulsations, the only characteristic spatial scale which can be discriminated in the system is the distance $z$ from the boundary. Hence, the turbulent viscosity can be determined only by 3 physical parameters: $z$, $v_\ast$ and $\rho$. The only combination of these parameters with the measurement units of kinematic viscosity is a product $v_\ast z$; and, as there in no dimensionless combinations of these parameters, according to the Buckingham $\pi$ theorem~\cite{Buckingham},
\begin{equation}
\nu_\mathrm{t}=\varkappa v_\ast z\,,
\label{eq-05}
\end{equation}
where $\varkappa$ is a dimensionless geometric factor, which can be determined empirically or from detailed numerical simulation of turbulent currents. It is known from experiments that $\varkappa\approx0.4$.

With imposed momentum flux (\ref{eq-04}) and inhomogeneous viscosity (\ref{eq-05}), one finds
\begin{equation}
\nu_\mathrm{t}\frac{\mathrm{d}u}{\mathrm{d}z}=\sigma_{xz}\,.
\label{eq-06}
\end{equation}
From (\ref{eq-06}) the velocity profile of the macroscopic average flow can be obtained
\begin{equation}
u=\frac{v_\ast}{\varkappa}\ln{\frac{v_\ast z}{\xi_0\nu}}\,,
\label{eq-07}
\end{equation}
where $\xi_0$ is the dimensionless integration constant. Expression~(\ref{eq-07}) is accurate at the spatial scale large compared to the scale where the molecular viscosity is of the same order of magnitude as the turbulent one, {\it i.e.}, for $z\gg l_0$, where $l_0=\nu/v_\ast$ is the thickness of so-called viscous sublayer. The shape of profile $u(z)$ for $z\sim l_0$ is known from experiments, and it is such that the condition of no slip along the boundary requires $\xi_0\approx 0.13$.

The case of flow along a nondeformable or weakly-deformable interface between two fluids is qualitatively similar to the case of flow along a rigid boundary, because the basic points used for derivation of eq.~(\ref{eq-05}) hold true. Although constants $\varkappa$ and $\xi_0$ may be different compared to the case of a rigid wall, laws (\ref{eq-05}) and (\ref{eq-07}) must hold valid.

Let us estimate the reference order of magnitude of parameters of the turbulent boundary layer for sensible situations with water flows, where $\nu=10^{-6}\mathrm{m^2/s}$, $v_\ast\sim0.01\,\mathrm{m/s}$, $z\sim0.1\,\mathrm{m}$. The thickness of the viscous sublayer, beyond which the average flow profile is accurately described by (\ref{eq-07}), is $l_0\sim10^{-4}\mathrm{m}$, the characteristic value of the argument of logarithm in (\ref{eq-07}) in the bulk is $\sim 10^4$ (which yields for the logarithm $\ln{10^4}\approx10$), and the characteristic average flow velocity~(\ref{eq-07}) $u\sim0.25\,\mathrm{m/s}$. For the stream canal of a common river, $v_\ast\sim0.05\,\mathrm{m/s}$ and depth $z\sim5\,\mathrm{m}$, one can estimate $l_0\sim2\cdot10^{-5}\mathrm{m}$, which yields the logarithm argument $\sim2.5\cdot10^6$ ($\ln{2.5\cdot10^6}\approx15$) and the maximal stream speed in the canal $u\sim2\,\mathrm{m/s}$.

\section{Solutions to Euler equation}
The solution to Euler equation~(\ref{eq-03}) with the constrain of the continuity equation~(\ref{eq-02}) can be sought in the potential form;
\begin{equation}
\vec{v}=-\nabla\varphi\,,
\label{eq-08}
\end{equation}
where $\varphi$ is the potential of the velocity field. Hence, eq.~(\ref{eq-02}) yields
\begin{equation}
\mathrm{\Delta}\varphi=0\,,
\label{eq-09}
\end{equation}
and Euler equation~(\ref{eq-03}) determines the pressure field in the fluid; $p=\rho\big(\varphi_t-(\nabla\varphi)^2/2-U(\vec{r})\big)$, where $U(\vec{r})$ is the potential of the mass force field, $\vec{g}=-\nabla{U(\vec{r})}$. The condition of impermeability of a nondeformable boundary in terms of velocity potential provides the boundary conditions for eq.~(\ref{eq-09}): $\partial\varphi/\partial n=0$.

Noteworthy, due to the harmonicity property~(\ref{eq-09}), potential flow~(\ref{eq-08}) satisfies also Navier--Stokes equation~(\ref{eq-01}). However, in the general case the boundary conditions for a viscous fluid cannot be satisfied with a potential flow; they have a higher order than the order which can be consistent with equation~(\ref{eq-09}).

\section{Low-viscosity flow}
\subsection{Equations and boundary conditions for average macroscopic flow}\label{sec41}
The average over turbulent pulsations fluid flow $\vec{u}(\vec{r},t)$ obeys the following equations with effective turbulent viscosity (see, {\it e.g.}, \cite{Schlichting}):
\begin{equation}
\rho\left(\frac{\partial\vec{u}}{\partial t}
 +(\vec{u}\cdot\nabla)\vec{u}\right)
 =-\nabla p+\nabla\cdot\hat{\sigma}_\mathrm{t}+\rho\vec{g}\,,
\label{eq-10}
\end{equation}
where effective turbulent viscosity stress tensor
\begin{equation}
\hat{\sigma}_\mathrm{t}(\vec{r},t)=\rho\,\nu_\mathrm{t}(\vec{r},t)
\left(\nabla\vec{u}+(\nabla\vec{u})^T\right)\,.
\label{eq-11}
\end{equation}
Here $p$ differs from the molecular pressure and contains a turbulent contribution; superscript $T$ indicates transposing. Strictly speaking, turbulent viscosity is not a scalar, but a tensor of a higher rank. However, for our treatment it is sufficient to use the approximation of scalar $\nu_\mathrm{t}$.

The turbulent viscosity property of primary importance for our consideration is that this viscosity tends to zero as one approaches a nondeformable boundary. This property is well seen in equation~(\ref{eq-05}). The persistence of this property for a broad range of situations can be substantiated as follows. By virtue of the fact that for realistic flows of water, for instance, the thickness of the viscous sublayer can be estimated as $l_0\sim10^{-6}-10^{-4}\mathrm{m}$, there is wide enough range of scales $l$ which are large compared to the viscous sublayer thickness, but small compared to the system scale $L\sim0.1-10\,\mathrm{m}$: $l_0\ll l\ll L$. At the distance of the order of magnitude of $l$ from the boundary of the flow domain, one can assume a spatially constant shear stress, neglect the boundary curvature and inhomogeneity of the average flow along the boundary, and the current to be statistically stationary at this spatial scale. These assumptions correspond to the basic points adopted for construction of the turbulent boundary layer theory in sect.~\ref{sec2} and, therefore, make results~(\ref{eq-05}) and (\ref{eq-07}) relevant at the scales we consider here. Hence, for $l_0\ll L$ the turbulent viscosity tends to zero as one approaches a nondeformable boundary.

The turbulent viscosity coefficient can also vanish at the free interface between two liquids which possesses a nonzero surface tension. The sufficient condition for this vanishing is the smallness of $l_0$ against the background of the interface curvature radius $r_0$: $l_0\ll r_0$. In this case, one can choose a scale $l$ such that $l_0\ll l\ll r_0$; at this scale, the interface is practically inflexible, {\it i.e.}, the result~(\ref{eq-05}) is valid, meaning vanishing turbulent viscosity at the interface. The value of $r_0$ for typical problems can be assessed from comparison between the gravity force and the surface tension force; $\rho r_0^3g\sim\tau r_0$, where $\tau$ is the surface tension coefficient. Hence, $r_0\sim\sqrt{\tau/(\rho g)}$. For water, one finds $r_0\sim3\,\mathrm{mm}$ and, for instance, with $l_0\sim10^{-5}-10^{-4}\mathrm{m}$ there is a wide range of possible values of $l$ satisfying condition $l_0\ll l\ll r_0$. The existence of this range means that the turbulent viscosity vanishes at such an interface.

\subsection{Solution to equations with turbulent viscosity}
With the turbulent viscosity vanishing at the boundary, one can seek the solution to equation~(\ref{eq-10}) in a potential form, as the problem is free from the no-slip boundary condition. For
\begin{equation}
\vec{u}=-\nabla\phi
\label{eq-12}
\end{equation}
the divergence of the turbulent viscosity stress tensor~(\ref{eq-11}) reads
\[
\nabla\cdot\hat{\sigma}_\mathrm{t}=-2\rho\nabla\nu_\mathrm{t}(\vec{r},t)\cdot\nabla\nabla\phi
 -2\rho\,\nu_\mathrm{t}(\vec{r},t)\nabla\mathrm{\Delta}\phi\,.
\]
Due to harmonicity of potential flow,
\[
\nabla\cdot\hat{\sigma}_\mathrm{t}=-2\rho\nabla\nu_\mathrm{t}(\vec{r},t)\cdot\nabla\nabla\phi\,.
\]
Employing the results for the turbulent boundary layer provided in sect.~\ref{sec2}, one can compare the order of magnitude of terms $\rho(\vec{u}\cdot\nabla)\vec{u}$ and $\nabla\cdot\hat{\sigma}_\mathrm{t}$ in eq.~(\ref{eq-10}) for potential flow~(\ref{eq-12}).

Indeed, according to (\ref{eq-05}),
\[
|\nabla\nu_\mathrm{t}(\vec{r},t)|\sim\varkappa\,v_\ast\,.
\]
Hence,
\[
\nabla\cdot\hat{\sigma}_\mathrm{t}
 \sim-\rho\varkappa\,v_\ast\nabla\nabla\phi
 \sim\rho\varkappa v_\ast\nabla u\,.
\]
The advective term (for a potential flow)
\[
\rho(\vec{u}\cdot\nabla)\vec{u}\sim\rho u\nabla u\,.
\]
The ratio of characteristic values of these two terms is
\[
\frac{|\nabla\cdot\hat{\sigma}_\mathrm{t}|}{|\rho(\vec{u}\cdot\nabla)\vec{u}|}
\sim\frac{\varkappa v_\ast}{u}\,,
\]
which is a small value at low viscosity, as we will show below in the text. Thus, in equation~(\ref{eq-10}) the term which makes it different from the Euler equation turns out to be small compared to the other terms.

The ratio $v_\ast/u$ is expected to be small at low viscosity; nonetheless, its characteristic value is yet to be defined. For an ideal fluid flow, which approximately corresponds to equation~(\ref{eq-10}) at low viscosity, the characteristic flow velocity near the boundary is of the same order of magnitude as the one in the bulk. Simultaneously, next to the boundary the logarithmic profile~(\ref{eq-07}) occurs, which can be employed for the assessment of the ratio
\begin{equation}
\frac{|\nabla\cdot\hat{\sigma}_\mathrm{t}|}{|\rho(\vec{u}\cdot\nabla)\vec{u}|}
\sim\frac{\varkappa v_\ast}{u}
\sim\frac{\displaystyle\varkappa^2}
 {\displaystyle\ln\frac{v_\ast L}{\xi_0\nu}}\,.
\label{eq-13}
\end{equation}

In the limit of arbitrary small viscosity the logarithm argument tends to infinity and, therefore, ratio~(\ref{eq-13}) tends to zero. This fact, indeed, allows one to neglect the term $\nabla\cdot\hat{\sigma}_\mathrm{t}$ against the background of $\rho(\vec{u}\cdot\nabla)\vec{u}$ when considering a potential flow~(\ref{eq-12}). Moreover, with the turbulent mechanism of eddy viscosity a potential flow becomes compatible with the physically natural boundary conditions (in contradistinction to the case of homogeneous viscosity, where a potential flow is admitted by the Navier-Stokes equation, but is generally incompatible with boundary conditions).

Thus, with a given geometry of the flow domain, the Euler equation with the boundary conditions for an inviscid fluid turns out to correspond the average over turbulent pulsations flow of a fluid of arbitrary small viscosity. In this sense, one can speak of the existence of the passage to the limit of inviscid fluid, and the Euler equation can be correctly employed for this limiting case.

\subsection{Small parameter characterising the accuracy of Euler equation}
The small parameter quantifying the accuracy of Euler equation for low-viscosity flows is remarkable. According to (\ref{eq-13}), this parameter is
\begin{equation}
\varepsilon=\frac{\displaystyle\varkappa^2}
 {\displaystyle\ln\frac{v_\ast L}{\xi_0\nu}}\,.
\label{eq-14}
\end{equation}
Parameter $\varepsilon$ is logarithmically small with respect to viscosity $\nu$ and the thickness of the viscous sublayer. Let us make two reference estimates for water flows ($\nu=10^{-6}\mathrm{m^2/s}$):
\\
1)~$v_\ast\sim0.01\,\mathrm{m/s}$, $L\sim0.1\,\mathrm{m}$ (typical for desktop installations). The logarithm argument is $\sim10^4$ and parameter $\varepsilon\approx0.02$.
\\
2)~$v_\ast\sim0.05\,\mathrm{m/s}$, $L\sim5\,\mathrm{m}$ (stream canal of a common river). The logarithm argument is $\sim2\cdot10^6$ and parameter $\varepsilon\approx0.01$.
\\
One can see that parameter $\varepsilon$ is quite small for these cases. The smallness of parameter $\varepsilon$ is logarithmically weak with respect to the ratio of the viscous sublayer thickness to the geometric size of the system; for the above estimates the decrease of $(l_0/L)$ by a factor $200$ resulted in the decrease of parameter $\varepsilon$ merely by a factor $2$.

\subsection{Mathematical description of low-viscosity multiphase flows}
Let us summarise what should be the equations and boundary conditions for flows of multiphase systems with a free interface at low viscosity.

In sect.~\ref{sec41}, it has been shown, that for an interface with nonzero surface tension the turbulent viscosity should vanish at the interface if $\nu/v_\ast\ll\sqrt{\tau/(\rho g)}$. Hence, at interfaces in multiphase systems the normal components of the current velocity in contacting phases should be matched, while tangential components can be arbitrary---the mutual slipping of flows is possible.

Within each phase the flow must be potential and smooth. There is an important point to be emphasised here: the Euler equation formally admits discontinuities of the velocity field and, in particular, mutual slipping of flows. However, the presence of nonzero turbulent viscosity in the bulk of each phase makes the flow discontinuities in the bulk impossible. The Euler equation is valid for mathematical description of the macroscopic average flow as long as this flow is a potential one and its potential in smooth within the given phase.

Besides the multiphase systems with nonzero surface tension, there are systems where interfaces do not possess surface tension. An example of such a system is in the case of contact between two mutually soluble liquids. In this case the mutual dissolution is operated by molecular diffusion and phases can remain well separated at time scales which are large compared to the hydrodynamic time scale. This conclusion holds valid for the system of two volumes of the same liquid but with different concentration of solute or fine suspension. An external force field ({\it e.g.}, gravity) can drive such a multiphase system to a stratified state, where the interface will possess certain effective resilience to deformation; being distorted, it will tend to restore a flat horizontal shape. The effective interface resilience can hinder the penetration of vortex motions through the interface and diminish the efficiency of the turbulent eddy transport of the average momentum. As a result, one should expect decrease of the turbulent viscosity at this interface. One can estimate the reference length scale $l_\ast$ of the interface deformations due to turbulent pulsations; $\rho v_\ast^2/2\sim\mathit{\Delta}\rho\,gl_\ast$, where $\rho$ is the characteristic density of liquids, $\mathit{\Delta}\rho$ is the density difference between two liquids. Whence,
\begin{equation}
l_\ast\sim\frac{\rho\,v_\ast^2}{\mathit{\Delta}\rho\,g}\,.
\label{eq-15}
\end{equation}
At the scales which are large compared to $l_\ast$, the basic points of sect.~\ref{sec2}, leading to expression~(\ref{eq-05}), are relevant, although the value of multiplier $\varkappa$ specific to this case should be determined empirically, not from a generalised analysis, and can differ from $0.4$. At these scales, it is natural to expect the turbulent viscosity to practically vanish near the interface.

\begin{table}[t]
\caption{Reference scale~(\ref{eq-15}) of perturbations of an interface with zero surface tension}
\center{
\begin{tabular}{lcc}
\hline
\hline
 \rule{0pt}{10pt} &
  $v_\ast=0.01\,\mathrm{m/s}$ &
  $v_\ast=0.05\,\mathrm{m/s}$ \\[5pt]
\hline
\rule{0pt}{15pt}
 $\mathit{\Delta}\rho/\rho=0.2$ & $5\cdot10^{-5}\mathrm{m}$ & $10^{-3}\mathrm{m}$ \\[5pt]
\rule{0pt}{15pt}
 $\mathit{\Delta}\rho/\rho=5\cdot10^{-4}$ & $0.02\,\mathrm{m}$ & $0.5\,\mathrm{m}$ \\[5pt]
\hline
\hline
\end{tabular}
}
\label{tab1}
\end{table}

In table \ref{tab1}, the values of $l_\ast$ are provided for the cases of desktop installations and stream canals of common rivers for a pair of mutually soluble heterogeneous liquids ($\mathit{\Delta}\rho/\rho\sim0.2$) and confluence of two fresh-water rivers with different natural water hardness ($\mathit{\Delta}\rho/\rho\sim5\cdot10^{-4}$). As an example of the latter, the confluence of the rivers Chusovaya and Sylva near the Ural Mountains, for which the detailed measurements on water properties and a persistent stratified state are available~\cite{Lyubimova-etal-2014}, can be mentioned. Here $\mathit{\Delta}\rho/\rho\sim3\cdot10^{-4}$ and during the winter period, when two rivers are covered by ice sheet, a stratified state with a two-layer flow is observed downstream from the confluence site and even upstream from it; the waters of the River Chusovaya overlay the water of the River Sylva.

Note, for the interface of the mixing layer, our analysis and conclusions are not applicable. Such an interface possesses neither surface tension nor any kind of effective resilience; therefore, there is no mechanism for a significant decrease of the turbulent viscosity and thus persistent existence of the potential discontinuity cannot be expected.

\section{Conclusion}
The passage to the limit of inviscid fluid for the Navier--Stokes equation exists and it leads to the Euler equation. In this limit, the fluid flow is turbulent, and the passage is essentially related to the properties of the turbulent boundary layer; specifically, the property of the effective turbulent viscosity, which is nonuniform in space, to vanish at the rigid boundaries and the liquid--liquid interfaces. With turbulent viscosity vanishing at the boundary, one can adopt for the macroscopic average flow the same boundary conditions as for an ideal fluid; the no-slip boundary condition does not appear. Such boundary conditions can be satisfied with a potential flow, which allows one to seek for the solution in a potential form. Simultaneously, the relative contribution of the term related to  turbulent viscosity into the equation for the macroscopic average flow turns out to be logarithmically small with respect to the ratio of the viscous sublayer thickness $l_0$ to the geometric size of the system $L$; this contribution is quantified by a small dimensionless parameter $\varepsilon=\varkappa^2/\ln[L/(\xi_0l_0)]$, where for a rigid boundary $\varkappa=0.4$ and $\xi_0=0.13$ (for a free interface, geometric coefficients $\varkappa$ and $\xi_0$ may be different). Thus, in the limit $l_0/L\to0$, the potential of the macroscopic average flow obeys the same equation to which the potential solutions to the Euler equation obey.

In this sense, one can speak of the existence of the passage to the Euler equation in the limit of arbitrary small viscosity; the solution to the Euler equation represents macroscopic fluid flow averaged over turbulent pulsations.


\section*{Acknowledgements}
The author is grateful to P.\ Frick for useful comments and discussion and acknowledges financial support by the Russian Science Foundation (Grant No.\ 14-21-00090).


\bibliographystyle{elsarticle-num}

\end{document}